\documentclass[12pt]{article}
\usepackage{amsmath}
\usepackage{amsfonts}
\usepackage{amssymb}
\usepackage{graphics}
\usepackage{graphicx}
\usepackage{amssymb}
\usepackage{cite}
\usepackage{setspace}
\usepackage{appendix}
\usepackage{graphicx} 

\usepackage{booktabs} 
\usepackage{array} 
\usepackage{paralist} 
\usepackage{verbatim} 
\usepackage{subfig} 

\usepackage{fancyhdr} 
\usepackage{sectsty}
\allsectionsfont{\sffamily\mdseries\upshape} 

\usepackage[nottoc,notlof,notlot]{tocbibind} 

\usepackage{amscd}
\usepackage{makeidx}
\newtheorem{teor}{Theorem}[section]

\newtheorem{defin}{Definition}[section]

\newcommand{\inv}{^{-1}}                                               
\newcommand{\beq}{\begin{equation}}
\newcommand{\eeq}{\end{equation}}
\newcommand{\bs}{\boldsymbol}
\numberwithin{equation}{section}

\title{Quantum Mechanics in non-inertial reference frames:  time-dependent rotations and loop prolongations}
\author{W.H.~Klink$^1$ and S.~Wickramasekara$^{1,2}$\\ 
$^1$ Department of Physics and Astronomy\\University of
Iowa\\
 Iowa City, IA 52242\\
$^2$ Department of Physics\\
Grinnell College\\ 
Grinnell, IA 50112\\}
\date{}
\begin{document}
\maketitle
\begin{abstract}
This is the fourth in a series of papers on developing a formulation of quantum mechanics in non-inertial reference frames. This formulation is grounded in a class of unitary cocycle representations of what we have called the Galilean line group, the generalization of the Galilei group to include transformations amongst non-inertial reference frames. These representations show that in quantum mechanics, just as the case  in classical mechanics,  the transformations to accelerating reference frames give rise to fictitious forces. In previous work, we have shown that there exist representations of the Galilean line group that uphold the non-relativistic equivalence principle as well as representations that violate the equivalence principle. In these previous studies, the focus was on linear accelerations. In this paper, we undertake an extension of the formulation to include rotational accelarations. We show that the incorporation of rotational accelerations requires a class of \emph{loop prolongations} of the Galilean line group and their unitary cocycle representations. We recover the  centrifugal and Coriolis force effects from these loop representations.  Loops are more general than groups in that their multiplication law need not be associative. Hence,  our broad theoretical claim is that a Galilean quantum theory that holds in arbitrary non-inertial reference frames requires going beyond groups and group representations, the well-stablished framework for implementing symmetry transformations in quantum mechanics. 
\end{abstract}

\section{Introduction}\label{sec1}
The purpose of this paper is to present to  a formulation of Galilean quantum mechanics in non-inertial reference frames, including both rotationally and linearly accelerating reference frames. In previous studies \cite{bk1,sw1,bksw}, we have introduced \emph{the Galilean line group}, the group of symmetry transformations amongst all accelerating reference frames in a Galilean spacetime, and propounded that Galilean quantum mechanics in non-inertial reference frames should be grounded in unitary cocycle representations of this group.  This is an extension of the well established claim that the standard quantum theory, a theory that holds in inertial reference frames, is grounded in unitary representations of the group of transformations of inertial reference frames in the relevant spacetime--the
 Poincar\'e group for the relativistic case and the Galilei group for the non-relativistic case. In particular, the unitary \emph{irreducible} representations of Poincar\'e and Galilei groups characterize elementary quantum systems, i.e., particles, in relativistic and non-relativistic quantum theory, respectively\footnote{Since there is a well-defined relativity in both cases, 
we will henceforth use the term \emph{Galilean quantum theory} in place of the common but inaccurate term  ``non-relativistic quantum theory".}. In both cases, mass and spin, the fundamental kinematic parameters that define an elementary particle, arise as eigenvalues of invariant operators in the enveloping algebra spanned by the generators of such unitary representations. 

The point of view we seek to advance is that in a Galilean quantum theory that holds in non-inertial reference frames, an elementary physical system is characterized by a unitary irreducible 
representation of the Galilean line group. If such a characterization is to be consistent with that of the established formulation of quantum mechanics in inertial reference frames,  the relevant representations of  the Galilean line group must be those which contain a unitary irreducible \emph{projective} representation of the Galilei group as a subrepresentation. 
Since projective unitary representations of a group are equivalent to true (vector) unitary 
representations of the central extensions of the group, we may recast this restriction on the admissible representations as an embedding criterion: 
construct a suitable extension of the Galilean line group that embeds a given central extension of the Galilei group. 
However, as shown in \cite{sw1}, while the Galilean line group naturally contains the Galilei group as a subgroup, it has  no central 
extensions  that contains a  central extension of the Galilei group. This motivates us to consider more general, non-central extensions of the line group. 
Even such general group extensions are possible, as also shown in \cite{sw1}, only when time-dependent rotations are omitted. 

The linear acceleration group, which consists of all linear acceleration transformations, time translations and constant rotations,  has non-central group extensions  
by the Abelian group of analytic functions on the real line. In fact, as shown in \cite{bksw}, the second cohomology group that defines these extensions is infinite 
dimensional. An important physical result that follows is that some elements of the second cohomology group, i.e., equivalence classes of cohomologous two-cocycles, lead to 
violations of the principle of equivalence in Galilean quantum mechanics, though in the classical limit these violations always disappear \cite{bksw}.  There also exist 
two-cocycles of the linear acceleration group which uphold the equivalence principle \cite{bk1,sw1}. 

The purpose of this paper is to extend the results of our previous work, which have been limited to linear accelerations, to include rotational accelerations. The key concept that we make use of is that of a 
\emph{loop}, a set closed under a composition rule that fulfills all the defining axioms of a group, except associativity and, as a result, the existence of a two-sided inverse. 
As such, loops are sometimes referred to as non-associative groups. 
While there exist no two-cocycles of the full Galilean line group (including time dependent rotations) taking values on the abelian group of real analytic functions and reducing to a two-cocycle of the Galilei group \cite{sw1,bksw}, we show here that there do exist three-cocycles fulfilling the reduction criterion and that these three cocycles lead to \emph{loop prolongations} of the Galilean line group which contain 
central extensions of the Galilei group. The main technical result we report in this paper is the construction 
of a certain class of representations of loop prolongations of the Galilean line group. These representations, constructed in section \ref{sec3}, 
 are defined by the following transformation formula for the generalized 
velocity eigenvectors $|\bs{q}\rangle$: 
\begin{equation}
\hat{U}^\times(\bar{g})\mid\bs{q}\rangle=e^{i\xi(\bar{g};\bs{q})}\mid\Lambda_{-b}\bs{q}'\rangle\label{1.1}
\end{equation}
where 
\begin{eqnarray}
\bar{g}&=&(\varphi,R,\bs{a},b)\nonumber\\
\xi(\bar{g};\bs{q})&=&m\varphi+m\left(\bs{q}'\cdot\bs{a}-\frac{1}{2}\bs{a}\cdot\dot{\bs{a}}+\frac{1}{2}(\Lambda_{-b}-1)\bs{q}'\cdot\bs{a}_{\bs{q}'}\right)-wb\nonumber\\
\bs{a}_{\bs{q}}&=&\int dt\,\bs{q},\quad \bs{q}=\frac{d}{dt}\bs{a}_{\bs{q}}\nonumber\\
\bs{q}'&=&R\bs{q}+\dot{R}\bs{a}_{\bs{q}}+\dot{\bs{a}}=\frac{d}{dt}(R\bs{a}_{\bs{q}}+\bs{a})=\frac{d}{dt}\bs{a}_{\bs{q'}}\nonumber\\
w&=&\text{invariant, internal energy}\label{1.2}
\end{eqnarray}
and $\Lambda_b$ is the shift operator, $(\Lambda_bf)(t)=f(t+b)$. Here, $R$ and $\bs{a}$ are, respectively, rotations and space translations with arbitrary time dependence. They furnish transformations to non-inertial reference frames with arbitrary accelerations. 
The parameter $b$ stands for time translations and  the set of elements $(R,\bs{a},b)$ constitute the Galilean line group. The $\varphi$ belongs to the Abelian group of real scalar functions of time, which provides the 
loop prolongation of the Galilean line group. As stated above, this loop prolongation is necessary for the cocycle representation \eqref{1.1} to reduce to a projective representation of the Galilei group, 
our primary physical requirement that limits the class of allowed representations. 

A noteworthy feature of the construction is that associativity is restored for the operators \eqref{1.1}, 
 as expected for linear operators in a vector space, despite the violation of the property in the loop prolongation $\{(\varphi,R, \bs{a},b) \}$. 
This result is based on the particular cocycle structure of \eqref{1.1}, where a 
three coboundary that appears in the triple product of operators exactly cancels the action of operators that represent the loop associators. The associators, defined below, are those loop elements that measure deviations 
from associativity, rather like commutators furnish a measure of non-commutativity. Alternatively, omitting $\varphi$, we may view the representations \eqref{1.1}  as representations 
of the Galilean line group with an intricate cocycle structure so that they do not correspond to vector representations of a group extension of the Galilean line group. 
From the results reported in \cite{bk1,sw1,bksw} and what is developed 
below in this paper, we note these representations completely describe the effects of fictitious forces due to both linear and rotational accelerations of reference frames. 
Consequently, our main proposition is that Galilean quantum mechanics in accelerating reference frames, both linearly and rotationally,  is grounded in the representations \eqref{1.1}.

The organization of the paper is as follows. In section \ref{sec2}, we provide a review of the Galilean line group, its group extensions and loop prolongations. 
In section \ref{sec3}, we construct a class of representations of loop prolongations of the Galilean line group by generalizing the method of induced representations, 
study their cocycle structure and derive their generators. 
In particular, we show that the Hamiltonian includes the effects of Coriolis and centrifugal forces in rotating reference frames.  In section \ref{sec4}, we will make a few concluding remarks. 
We will collect a few important facts about algebraic cohomology of groups and loops  in Appendix \ref{appendix}.

\section{Galilean line group and its loop prolongations}\label{sec2}
The Galilei group can be defined as a group of transformations on the spacetime manifold: 
 \begin{eqnarray}
\bs{x}'&=&R\bs{x}+\bs{v}t+\bs{a}\nonumber\\
t'&=&t+b,\label{2.1}
\end{eqnarray}
where $R$, $\bs{v}$, $\bs{a}$  and $b$ are, respectively, rotations, velocity boosts, space translations and time translations. Thus, the Galilei group transforms inertial reference frames 
amongst themselves. Unlike Lorentz boosts, 
Galilean boosts do not affect the time coordinate of the manifold and their action on the space coordinates, $\bs{v}t$, is simply a translation that depends on time linearly. 
Therefore, we can write  \eqref{2.1} more compactly as 
\begin{eqnarray}
\bs{x}'&=&R\bs{x}+\bs{a}(t)\nonumber\\
t'&=&t+b\label{2.2}
\end{eqnarray}
where $\bs{a}(t)=\bs{a}+\bs{v}t$. This structure suggests a generalization of $\bs{a}(t)$ to include space translations with arbitrary time dependence, such as $\bs{a}(t)=\frac{1}{2}\bs{a}t^2$, 
 as a way of transforming among 
accelerating reference frames. Likewise, we can accommodate rotational accelerations by letting the 
rotation matrices $R$ be functions of time. Therewith, we have the transformations 
\begin{eqnarray}
\bs{x}'&=&R(t)\bs{x}+\bs{a}(t)\nonumber\\
t'&=&t+b\label{2.3}
\end{eqnarray}
For the sake of notational economy, let us denote the transformations \eqref{2.3} simply by $\{(R,\bs{a},b)\}$ with the understanding that $R$ and $\bs{a}$ are now functions of time 
that get evaluated at the time coordinate of the spacetime point on which they act.  By the successive operation of two elements on a given spacetime point $(\bs{x},t)$, we obtain 
a composition rule for the set $\{(R,\bs{a},b)\}$:
\begin{equation}
(R_2, \bs{a}_2, b_2)(R_1, \bs{a}_1, b_1) = ((\Lambda_{b_1} R_2) R_1 , \ (\Lambda_{b_1} R_2)\bs{a}_1 + \Lambda_{b_1} \bs{a}_2 ,\  b_2 + b_1)\label{2.5}.
\end{equation}
where $\Lambda_b$ is the shift operator $(\Lambda_bf)(t)=f(t+b)$ and the products of time dependent $R$ and $\bs{a}$ stand for point-wise multiplication, e.g., 
$(\Lambda_{b_1}R_2)R_1(t):=(\Lambda_{b_1}R_2)(t)R_1(t)=R_2(t+b_1)R_1(t)$. The element $(I,\bs{0},0)$ is the identity element for \eqref{2.5}. 

The associativity of the composition  rule \eqref{2.5} follows from that fact that $b\to \Lambda_b$ is a homomorphism. 
Also, each element $(R,\bs{a},b)$ has an inverse under \eqref{2.5}:
\begin{equation}
(R, \bs{a}, b)\inv = (\Lambda_{-b} R\inv, -\Lambda _{-b}(R\inv \bs{a}) , -b).\label{2.6}
\end{equation}
Therefore, the set of time dependent rotations and space translations, along with time translations, form an infinite dimensional group under \eqref{2.5}. In \cite{sw1}, this group was called the Galilean line group. 
We denote it here by ${\cal G}$. 

Note that time dependent rotations ${\cal R}:=\{(R,\bs{0},0)\}$,  time dependent spatial translations ${\cal A}:=\{(I,\bs{a},0)\}$ and time translations ${\cal B}:=\{(I,0,b)\}$ are all 
subgroups of $\cal G$.  The subgroup ${\cal E}(3):=\{(R,\bs{a},0)\}$, which is 
the group of maps from the real line to the Euclidean group in three dimensions, was called the Euclidean line group in \cite{bk1}. 
It describes all transformations between linearly and rotationally accelerating reference frames. The Euclidean line group is the semidirect product of 
${\cal A}$ and ${\cal R}$: 
\begin{equation}
\mathcal{E}(3)=\mathcal{A}\rtimes\mathcal{R}.\label{2.7}
\end{equation}
Furthermore, from \eqref{2.5} we note that each $(I,\bs{0},b)\in{\cal B}$ induces an automorphism on the subgroup ${\cal E}(3)$, given by $\Lambda_b$. 
Therefore,  
\begin{equation}
\mathcal{G}  = {\mathcal E}(3)\rtimes_{\Lambda}\mathcal{B}= ( \mathcal{A} \rtimes \mathcal{R} ) \rtimes  _{\Lambda} \mathcal{B}.\label{2.8}
\end{equation}
That is, the Galilean line group is the semidirect product of the time translation subgroup and the line group of the three dimensional Euclidean group\footnote{\label{footnote1}{Note 
that our use of the external semidirect product differs slightly from the canonical definition.
 Looking at the subgroup of space and time translations, we have $(a_2, b_2)(a_1, b_1) = (\Lambda_{b_1}a_2 + a_1, b_2 + b_1)$, whereas a connonical external semi-direct product would 
 give $(a_2, b_2)(a_1, b_1) = (a_2 + \Lambda_{b_2}a_1, b_2 + b_1)$.     
Had we chosen our notation so that multiple transformations composed from left to right instead of right to left, then the stucture of the Galilei Galilean Line Group would look exactly like that of a semidirect product. That is to say, our semidirect product definition is the dual of the more standard one.}}. 

The subset $\{(R^0,\bs{a},b)\}$ of $\cal G$ consisting of constant rotations $R^0$, arbitrary space translations and time translations is also a subgroup of ${\cal G}$. We call this subgroup the linear acceleration 
group and denote it by ${\cal A}_{linear}=\{(R^0,\bs{a},b)\}$. Here and below, we denote quantities that are constant functions of time by the superscript $0$.

As mentioned in the Introduction, a physically meaningful representation of $\cal G$ must contain as a subrepresentation a unitary projective representation of the Galilei group. Since a projective representation of a group is equivalent to a true representation of a central extension of the group, we expect the physically meaningful representations of ${\cal G}$ to correspond to representations of a group extension of $\cal G$ so that this extension contains a given central extension of the Galilei group. As shown in \cite{sw1}, there exist no central extensions of $\cal G$ in which a given central extension of the Galilei group may be embedded. However, as also shown in \cite{sw1}, when time dependent rotations are excluded so that the symmetry group is ${\cal A}_{linear}$,  there exist group extensions fulfilling  the embedding criterion, although these extensions are non-central.  In fact, as shown in \cite{bksw}, there exist infinitely many different equivalence classes of cohomologous two-cocycles that define extensions of ${\cal A}_{linear}$ that fulfill the embedding condition. Of this infinity of inequivalent two-cocycles, for the sake of definiteness and simplicity, we will limit the discussion in this paper to the following two-cocycle,
\begin{equation}
\omega_2(g_2,g_1)=\frac{1}{2}\left(\left( \Lambda_{b_1}\bs{a}_2\right)\cdot R^0_2\dot{\bs{a}}_1-\left(\Lambda_{b_1}\dot{\bs{a}}_2\right)\cdot R^0_2\bs{a}_1\right), \label{2.9}
\end{equation}
and its generalization \eqref{2.9b} to include time dependent rotations. 

The two-cocycle \eqref{2.9} gives rise to a group extension $\tilde{\cal A}_{linear}=\{(\varphi,g): \ g\in{\cal A}_{linear}\}$ of ${\cal A}_{linear}$ by the Abelian group ${\cal F}({\mathbb{R}})$ of analytic real scalar functions 
$\varphi$ on $\mathbb{R}$. The composition rule under which $\tilde{\cal A}_{linear}$ becomes a group is
\begin{equation}
(\varphi_2,g_2)(\varphi_1,g_1)=(\Lambda_{b_1}\varphi_2+\varphi_1+\omega_2(g_2,g_1),g_2g_1)\label{2.10}
\end{equation}
The automorphism $\Lambda_b$ makes the extension subgroup $(\varphi,e)$ non-commutative with the time translation 
subgroup $(I,\bs{0},b)$.  Therefore, $\tilde{\cal A}_{linear}$ is a non-central extension of ${\cal A}_{linear}$. 

It is straightforward to verify that the function \eqref{2.9} satisfies the two-cocycle condition \eqref{3.10} (with $\sigma(g)=\Lambda_b$): 
\begin{equation}
(\delta_2\omega_2)(g_3,g_2,g_1)=\Lambda_{b_1}\omega_2(g_3,g_2)+\omega_2(g_3g_2,g_1)-\omega_2(g_2,g_1)-\omega_2(g_3,g_2g_1)=0.\label{2.11}
\end{equation}
This property, in turn, ensures that the composition rule \eqref{2.10} is associative. 
Further, when acceleration transformations are restricted to be Galilean transformations, $\bs{a}(t)=\bs{a}^0+\bs{v}t,$
the function \eqref{2.9} reduces to a two-cocycle of the Galilei group:
\begin{equation}
\omega_2(g_2,g_1)=\frac{1}{2}\left(\bs{a}_2^{(0)}\cdot R^0_2\bs{v}_1-\bs{v}_2\cdot R^0_2\bs{a}_1^{(0)}+b_1\bs{v}_2\cdot R^0_2\bs{v}_1\right)\label{2.12}
\end{equation}
This shows that the extension $\tilde{\cal A}_{linear}$ contains a central extension of the Galilei group.

In \eqref{2.9}-\eqref{2.11}, while the spatial translations may be arbitrary functions of time, rotations are required to be time independent. 
The main mathematical difficulty with time dependent rotations is apparent from the structure of the two-cocycle \eqref{2.9}. Since \eqref{2.12} (or, any other 
function that differs from it by a two-coboundary) involves Galilean boosts $\bs{v}$, any two-cocycle of $\cal G$ with the correct reduction to a Galilean two-cocycle 
must involve derivatives of space translations. However, when rotations are also time dependent, 
time dependent space translations $\bs{a}$ and their derivatives $\dot{\bs{a}}$ do not transform the same  way under rotations owing to the inhomogeneous term in the derivative
$\frac{d}{dt}\left(R(t)\bs{a}(t)\right)=R(t)\dot{\bs{a}}(t)+\dot{R}(t)\bs{a}(t)$.  This situation is entirely analogous to the way the derivatives of fields transform under gauge transformations. In the current setting, 
the trouble is algebraic: the term $\dot{R}(t)\bs{a}$ leads to a violation of the two-cocycle condition \eqref{2.11}. For instance, 
if we let rotations be time dependent, the two-cochain of \eqref{2.9} reads\footnote{Our notation is that, unless separated by parentheses, the 
 automorphism $\Lambda_b$ acts only on the group element immediately to the right of it. 
Thus, $\Lambda_{b_1}R_2\bs{a}_1\equiv\left(\Lambda_{b_1}R_2\right)\bs{a}_1$.}
\begin{equation}
\omega_2(g_2,g_1)=\frac{1}{2}\left(\Lambda_{b_1}\bs{a}_2\cdot\Lambda_{b_1}R_2\dot{\bs{a}_1}-\Lambda_{b_1}\dot{\bs{a}_2}\cdot\Lambda_{b_1}R_2\bs{a}_1\right)\label{2.9b}
\end{equation}
and the coboundary operator on \eqref{2.9b} yields:
\begin{eqnarray}
(\delta_2\omega_2)(g_3,g_2,g_1)&=&\frac{1}{2}\Lambda_{b_2+b_1}\dot{R_3}\Lambda_{b_1}\bs{a}_2\cdot\Lambda_{b_2+b_1}R_3\Lambda_{b_1}R_2\bs{a}_1\nonumber\\
&&\quad +\frac{1}{2}\Lambda_{b_2+b_1}\bs{a}_3\cdot\Lambda_{b_2+b_1}R_3\Lambda_{b_1}\dot{R}_2\bs{a}_1\nonumber\\
&=&\frac{1}{2}\Lambda_{b_1}\Omega_2\cdot(\Lambda_{b_1}R_2\bs{a}_1\times\Lambda_{b_2+b_1}(R_3^T\bs{a}_3))\nonumber\\
&&\qquad-\frac{1}{2}\Lambda_{b_2+b_1}\Omega_3\cdot(\Lambda_{b_1}\bs{a}_2\times\Lambda_{b_1}R_2\bs{a}_1)\not=0\label{2.13}
\end{eqnarray}
where the angular velocity vector $\Omega$ is related to the time derivative of the rotation matrix by the usual formula $\Omega\times\bs{a}=\dot{R}R^T\bs{a}$. The  expression \eqref{2.13} is to be compared with \eqref{2.11}. 

For every function $\omega(g_2,g_1)$ on $G\times G$, where $G$ is a group, that takes values in an Abelian group $A$ and  fulfills the two-cocycle condition \eqref{3.10}, there exists a group extension of $G$ by $A$. 
\emph{In particular, an associative composition rule can be defined so that this extension of $G$ by $A$ is again a group if and only if $\omega(g_2,g_1)$ fulfills the two-cocycle condition \eqref{3.10}}.
Therefore, \eqref{2.13} tells us that the two-cochain \eqref{2.9} does not lead to a group extension of the Galilean line group. This difficulty persists for any two-cochain that involves the derivatives of spatial 
translations, such as the more general ones considered in \cite{bksw}:
\begin{equation}
\omega_2(g_2,g_1)=\frac{1}{2}\Lambda_{b_1}\bs{B}(\bs{a}_2)\cdot\Lambda_{b_1}R_2\bs{C}(\bs{a}_1)-\frac{1}{2}\Lambda_{b_1}\bs{C}(\bs{a}_2)\cdot\Lambda_{b_1}R_2\bs{B}(\bs{a}_1)\label{2.14}
\end{equation}
where, 
\begin{eqnarray}
\bs{B}(\bs{a})&=&\sum_{n=0}^\infty \beta_n\frac{d^n\bs{a}}{dt^n}\nonumber\\
\bs{C}(\bs{a})&=&\sum_{n=0}^\infty \gamma_n\frac{d^n\bs{a}}{dt^n}.\label{2.15}
\end{eqnarray}
and $\beta_n$ and $\gamma_n$ are arbitrary constants. 

\subsection{Loops and loop prolongations of groups}\label{sec2.2}
Since the two-cocycle condition \eqref{2.11} fails for any two-cochain  that involves derivatives, we may at best consider an  extension of ${\cal G}$ by the group of functions $\varphi\in{\cal F}({\mathbb{R}})$ with a \emph{non-associative} composition rule. As we will shortly see, the resulting structure is a \emph{loop}. Since loops are not widely used in physics, here we collect some basic definitions and facts, mainly extracted from \cite{em}, before 
constructing the loop prolongation of the Galilean line group in the next subsection. 

{\defin{A \emph{loop} $L$ is a set with a composition rule $L\times L\to L$, called multiplication and denoted by $(a,b)\to ab$, such that the following axioms hold:}
\begin{enumerate}
\item For  $a, b\in L$,\ \text{the element} $ab\in L$ \text{is uniquely defined.}
\item \text{There is an element} $1\in L$, called \emph{identity},  \text{such that} $a1=1a=a$ \text{for all}\ $a\in L$. 
\item For any $a,b\in L,$\  {there exist unique elements} $x,y\in L$ \text{that solve the equations} $ax=b$\ \text{and}\ $ya=b$.
\end{enumerate}\label{def2.2.1}}
Note that, unlike for a group, associativity is not required. From the last requirement of the definition, for any $a,b,c\in L$, there exists an element 
$A(a,b,c)\in L$, called an \emph{associator}, such that
\begin{equation}
a(bc)=A(a,b,c)[(ab)c]\label{2.2.1}
\end{equation}
Thus, associators provide a measure of the lack of associativity, much like commutators register deviations from commutativity. 

A loop can be obtained by putting two groups together, a construction rather similar to that of group extensions. The relevant notion is a loop prolongation of a group $G$ by an Abelian group $K$: 
{\defin{Let $G$ be a group, $K$ an Abelian group and $\sigma(g)$ an automorphism on $K$ for every $g\in G$. Then, the pair $(L,\chi)$ is called a \emph{loop prolongation} 
of $G$ by $K$ if the following axioms hold:}
\begin{enumerate}
\item $L$ is a loop such that (the canonical injection of) $K$ is a subgroup of $L$. 
\item $\chi$ is a homomorphism of $L$ onto $G$ with kernel $S$ containing ${K}$. 
\item $A(s,a,b)=A(a,s,b)=1$ for all $s\in S$, $a,b,\in L$. 
\item $A(a,b,c)s=sA(a,b,c)$ for all $s\in S$,  $a,b,c \in L$. 
\item $a\circ{k}={\sigma(\chi(a)^{-1})k}$, where $a\circ k$ is the \emph{inner transformation} defined by \eqref{2.2.2}, 
 for all $a\in L$ and $k\in K$. 
\end{enumerate}\label{def2.2.2}}
The third and fourth conditions put limits the non associativity of $L$. The third condition shows that a triple product in which the first or second element belongs to 
$S$ is associative.  Further, as seen from \eqref{2.2.6}, these axioms also ensure that associativity holds for any triple product in which the first or second element is an associator.  

Further properties of loops and loop prolongations, including the statement of the main existence theorem of Eilenberg and MacLane \cite{em}, are discussed in Appendix \ref{appendix}.

\subsection{Galilean line loop}\label{sec2.3}
The function \eqref{2.9b} is a two-cochain on the Galilean line group ${\cal G}$ taking values on ${\cal F}(\mathbb{R})$, the Abelian group of real analytic scalar functions. For notational economy, 
let us denote the latter group simply by ${\cal F}$. For $g\in{\cal G}$, $\sigma(g)=\Lambda_b$ is 
an automorphism on ${\cal F}$. Therefore, by Theorem \ref{theorem2.2.1}, there exists a loop prolongation $(\bar{\cal G},\chi)$ of ${\cal G}$ by ${\cal F}$. We refer to $(\bar{\cal G},\chi)$ as 
the \emph{Galilean line loop}. 

The construction of $(\bar{\cal G},\chi)$ is straightforward. To that end, let $\bar{G}:=\{\bar{g}=(\varphi,g):\ \varphi\in{\cal F}, g\in{\cal G}\}$ with the composition rule
\begin{equation}
(\varphi_2,g_2)(\varphi_1,g_1)=(\Lambda_{b_1}\varphi_2+\varphi_1+\omega_2(g_2,g_1), g_2g_1)\label{2.3.1}
\end{equation}
where $\omega_2(g_2,g_1)$ is defined by \eqref{2.9b}. 
Let $\chi:\ \bar{\cal G}\to{\cal G}$  be defined by  $\chi(\bar{g})=g$, i.e., 
\begin{equation}
\chi(\bar{g})\equiv\chi(\varphi,g)\equiv\chi(\varphi,R,\bs{a},b)=g=(R,\bs{a},b)\label{2.3.2}
\end{equation}
We must verify that $(\bar{\cal G},\chi)$ fulfills the axioms of Definition \ref{def2.2.2}. 
\begin{enumerate}
\item ${\bar{G}}$ is a loop. To show this, we verify the conditions of Definition \ref{def2.2.1}: 
\begin{enumerate}
\item For any $(\varphi_1,g_1),\ (\varphi_2,g_2)\in\bar{\cal G}$, the composition $(\varphi_1,g_1)(\varphi_2,g_2)$ is well-defined by \eqref{2.3.1} and \eqref{2.9b} as an element of $\bar{\cal G}$. 
\item The element $(0,e)\in{\cal G}$, where $e=(I,\bs{0},0)$ is the identity of $\cal G$, is the identity of $\bar{\cal G}$: 
\begin{equation}
(0,e)(\varphi,g)=(\varphi,g)=(\varphi,g)(0,e)
\end{equation}
\item For any $(\varphi_1,g_1),\ (\varphi_2,g_2)\in\bar{\cal G}$, the equations
\begin{eqnarray}
(\varphi_1,g_1)(\varphi,g)&=&(\varphi_2,g_2)\nonumber\\
(\tilde{\varphi},\tilde{g})(\varphi_1,g_1)&=&(\varphi_2,g_2)\label{2.3.3}
\end{eqnarray}
have unique solutions in $\bar{\cal G}$: 
\begin{eqnarray}
(\varphi,g)&=&(\varphi_2-\Lambda_{b_2-b_1}\varphi_1-\omega_2(g_1,g_1^{-1}g_2), g_1^{-1}g_2)\nonumber\\
(\tilde{\varphi},\tilde{g})&=&\left(\Lambda_{-b_1}\left(\varphi_2-\varphi_1-\omega_2(g_2g_1^{-1},g_1)\right), g_2g_1^{-1}\right)\label{2.3.4}
\end{eqnarray}
\end{enumerate}
We may identify $\cal F$ with its canonical embedding $\varphi\to (\varphi,e)$ in $\bar{\cal G}$. Then, since $\omega_2(g,e)=\omega_2(e,g)=0$, it follows from \eqref{2.3.1} that 
$\{(\varphi,e),\ \varphi\in{\cal F}\}$ is a subgroup of $\bar{\cal G}$. 
\item The mapping $\chi:\ \bar{\cal G}\to{\cal G}$ defined by \eqref{2.3.2} is a homomorphism, 
\begin{equation}
\chi\left(\bar{g}_2\bar{g}_1)\right)=g_2g_1=\chi(\bar{g}_2)\chi(\bar{g}_1),\label{2.3.5}
\end{equation}
onto $\cal G$. Its kernel is $\{(\varphi,e)\}$. Note that unlike in the general case covered by Definition \ref{def2.2.2} and discussed in Appendix \ref{appendix}, 
here the kernel of the homomorphism is precisely the group 
$\cal F$ by which $\cal G$ is prolongated. 
\item A direct calculation using \eqref{2.3.1} gives 
\begin{eqnarray}
\bar{g}_3\left(\bar{g}_2\bar{g}_1\right)&=&\left(\Lambda_{-(b_3+b_2+b_1)}(\delta_2\omega_2)(g_3,g_2,g_1), e\right)\left[(\bar{g}_3\bar{g}_2)\bar{g}_1\right]\nonumber\\
&=&A(\bar{g}_3,\bar{g}_2,\bar{g}_1)\left[(\bar{g}_3\bar{g}_2)\bar{g}_1\right]\label{2.3.6}
\end{eqnarray}
where $(\delta_2\omega_2)(g_3,g_2,g_1)$ is the two-coboundary \eqref{2.13}. Thus, associators
\begin{equation}
A(\bar{g}_3,\bar{g}_2,\bar{g}_1)=\left(\Lambda_{-(b_3+b_2+b_1)}(\delta_2\omega_2)(g_3,g_2,g_1), e\right)\label{2.3.7}
\end{equation}
in fact only depend on the projections $\chi(\bar{g}): \ A(\bar{g}_3,\bar{g}_2,\bar{g}_1)=A(g_3,g_2,g_1)$. 
Therefore, from the explicit form \eqref{2.13} of $\delta_2\omega_2$, we note that \eqref{2.3.7} vanishes whenever 
$\bar{g}_3$ or $\bar{g}_2$ (but \emph{not} $\bar{g}_1$) is of the form $(\varphi,e)$. 
\item From \eqref{2.3.7}, all associators lie in $\cal F$. Hence, they commute with all elements of $\cal F$, the kernel of $\chi$. 
\item Since for any $\bar{g}=(\varphi,g)$
\begin{equation}
(\varphi, g)(\psi,e)=(\varphi+\psi,g)=(\Lambda_{-b}\psi,e)(\varphi,g),\label{2.3.8}
\end{equation}
it follows from definition \eqref{2.2.2}, 
\begin{equation}
\bar{g}\circ(\psi,e)=(\Lambda_{-b}\psi,e)=(\sigma(\bar{g}^{-1})\psi,e)\label{2.3.9}
\end{equation}
\end{enumerate}
The above considerations show that $(\bar{\cal G},\chi)$ is a loop prolongation of $\cal G$ by $\cal F$. In the classification scheme of \cite{em}, $(\bar{\cal G},\chi)$ is of class $K_0^S$. 

In addition to the above defining axioms, there are some additional properties that $(\bar{\cal G},\chi)$ fulfills. Since they are rather straightforward to establish, we omit the proofs. 
\begin{enumerate}
\item Since the kernel of $\chi$ coincides with $\cal F$, from \eqref{2.2.6}, the associativity holds for any triple product that involves at least one element of $\cal F$. 
\item Each $\bar{g}$ has a two-sided inverse in $(\bar{\cal G},\chi)$: 
\begin{equation}
\bar{g}^{-1}=\left(-\Lambda_{-b}\varphi, g^{-1}\right)\label{2.3.10}
\end{equation}
\item The prolongation of the Euclidean line group by $\cal F$, ($\bar{{\cal E}}(3),\chi)$, where $\bar{\cal E}(3)=\{(\varphi,R,\bs{a},0)\},$ is a subloop of $(\bar{\cal G}, \chi)$. 
\item  Any subloop on which the cocycle condition \eqref{2.11} holds is a subgroup of $(\bar{\cal G},\chi)$. In particular, the subset consisting of constant rotations is a subgroup, the non-central group extension 
of ${\cal A}_{linear}$ introduced in \cite{sw1} and highlighted above in \eqref{2.10}. The following subsets are also subgroups of $(\bar{\cal G},\chi)$:
\begin{equation}
\{(\varphi,R,\bs{0},b)\},\quad\{(\varphi,I,\bs{0},b)\},\quad\{(0,I,\bs{0},b)\}\label{2.3.11}
\end{equation}
\item The subset $\{(\varphi, I, \bs{a},0)\}$ is  a normal subgroup (see Appendix \ref{appendix}) of $(\bar{\cal G},\chi)$:
\begin{eqnarray}
&&(\psi, I, \bs{\tt{a}},0)(\varphi,R,\bs{a},b)=\nonumber\\
&&\qquad(\varphi,R,\bs{a},b)(\Lambda_b\psi+\frac{1}{2}(\dot{\bs{a}}\cdot\Lambda_b\bs{\tt{a}}-\bs{a}\cdot\Lambda_b\dot{\bs{\tt{a}}})+\frac{1}{2}\Omega\cdot(\Lambda_b\bs{\tt{a}}\times\bs{a}),I,R^{-1}\Lambda_b\bs{\tt{a}},0)\nonumber\\
\label{2.3.12}
\end{eqnarray}
\item When rotations are constant and translations are of the form $\bs{a}(t)=\bs{a}_0+\bs{v}t$, the loop $(\bar{\cal G},\chi)$ reduces to a central group extension of the Galilei group. 
\end{enumerate}

\section{Unitary representations of the Galilean line loop}\label{sec3}
According to the philosophy articulated in the introduction, it is reasonable to attempt to ground a Galilean quantum theory  that holds in both rotationally and linearly accelerating reference frames in a suitable class of unitary representations of the Galilean line loop $(\bar{\cal G},\chi)$. Here we immediately encounter a diffuclty: since linear operators in a vector space form an associative algebra under the usual composition rule $(\hat{A}\hat{B})\phi=\hat{A}(\hat{B}\phi)$ (where $\phi$ is an arbitrary element of the common domain of the set of operators), it appears that it would be impossible to define a faithful homomorphism from a non-associative loop into the associative operator algebra in a Hilbert space. The way of out this difficulty rests on two key points. First, in quantum mechanics we must always allow for cocycle representations. Second, the associators of $(\bar{\cal G}, \chi)$ are determined by a three-coboundary on ${\cal G}$.  As we will show below, these two properties can be exploited to construct a representation that is consistent with the associativity of operator composition in a Hilbert space. 
The representation is constructed so that the subgroup $\cal F$, and thus also the set of all associators,  is represented by phase factors. The cocycle structure of the representation is such that the three-coboundary that results from a triple product of operators exactly cancels the corresponding associator. For the sake of simplicity, we will limit ourselves to zero spin representations, although the construction can be extended to higher spin cases in a rather natural way. 

\subsection{Review of representations of extended linear acceleration group}\label{sec3.1}
 The analysis of \cite{sw1} dealt with the extended linear acceleration group $\tilde{{\cal A}}_{linear}$ and showed 
that there exist unitary representations of this group that uphold the principle of equivalence. In \cite{bksw}, it was shown there also exist representations of the extended linear acceleration group that violate the equivalence principle. In both cases, the analysis was restricted to linear accelerations mainly because of the absence of a group extension of the full Galilean line group that embeds a given central extension of the Galilei group. 

The representations of the extended linear acceleration group were constructed in \cite{sw1,bksw} by the method of induced representations. The starting point of the construction is the observation that, when rotations are not time dependent, the constant space translations $\bs{a}^0$ and the constant functions $\varphi^0\in{\cal F}$ form an invariant abelian subgroup of the extended linear acceleration group. As is well known, the unitary irreducible representations of an Abelian group are one dimensional. From such a representation of this invariant Abelian subgroup, 
\begin{equation}
\hat{U}^\times(\varphi^0,I,\bs{a}^0,0)\mid\bs{q}\rangle=e^{im(\varphi^0+\bs{q}\cdot\bs{a}^0)}\mid\bs{q}\rangle,\label{3.1.1}
\end{equation}
we may induce a unitary irreducible representation of the whole extended linear acceleration group: 
\begin{equation}
\hat{U}^\times(\varphi,R^0,\bs{a},b)\mid\bs{q}\rangle=e^{im\varphi+i\xi(g,\bs{q})}\mid\bs{q}'\rangle\label{3.1.2}
\end{equation}
where 
\begin{eqnarray}
\bs{q}'&=&R^0\bs{q}+\dot{\bs{a}}\nonumber\\
\xi(g;\bs{q})&=&m\left(\bs{q}'\cdot\bs{a}-\frac{1}{2}\bs{a}\cdot\dot{\bs{a}}+\frac{1}{2}(\Lambda_{-b}-1)\bs{q}'\cdot\bs{a}_{\bs{q}'}\right)-wb\nonumber\\
\bs{a}_{\bs{q}}&=&\int_0^t dt\, \bs{q}\quad (\dot{\bs{a}}_{\bs{q}}=\bs{q})\nonumber\\
w&=&\text{invariant, internal enegry}\label{3.1.3}
\end{eqnarray}
The superscript $^\times$ in the operators $\hat{U}^\times(g)$ is to indicate that these operators are defined in the topological dual of a suitable test function space  which 
the generalized vectors $\mid\bs{q}\rangle$ inhabit. 
The unitary operators $\hat{U}(g)$ in the corresponding Hilbert space are related to $\hat{U}^\times(g)$ by the duality formula
\begin{equation}
(\hat{U}(g)\psi)(\bs{q}):=\langle \hat{U}(g)\psi\mid\bs{q}\rangle^*=\langle\psi\mid\hat{U}^\times(g^{-1})\bs{q}\rangle^*\label{3.1.4}
\end{equation}
where $^*$ denotes complex conjugation. 

From the first equality of \eqref{3.1.3}, we note that for a given velocity $\bs{q}$, there exists a space translation $\bs{a}_{\bs{q}}$ such that $\dot{\bs{a}}_{\bs{q}}=\bs{q}$ so that $\bs{a}_{\bs{q}}$ 
can be considered a boost that transforms the rest vector $\bs{0}$ to velocity $\bs{q}$.  The group element $g_{\bs{q}}\equiv(0,I,\bs{a}_{\bs{q}},0)$ was chosen in \cite{sw1} as the \emph{standard boost} and the generalized eigenvector 
$\mid\bs{q}\rangle$ was \emph{defined} as 
\begin{equation}
\mid\bs{q}\rangle:=e^{-\frac{i}{2}m\bs{q}\cdot\bs{a}_{\bs{q}}}U^\times(g_{\bs{q}})\mid\bs{0}\rangle\label{3.1.5}
\end{equation}

The cocycle structure of the representation can be deduced from the repeated application of \eqref{3.1.2}: 
\begin{equation}
\hat{U}^\times(g_2)\hat{U}^\times(g_1)\mid\bs{q}\rangle=e^{i\xi_2(g_2,g_1,\bs{q})}\hat{U}^\times(g_2g_1)\mid\bs{q}\rangle\label{3.1.6}
\end{equation}
where 
\begin{equation}
\xi_2(g_2,g_1,\bs{q})=(1-\Lambda_{b_1})\varphi_2+(\Lambda_{-b_1}-1)\omega_2(g_2,g_1g_{\bs{q}})\label{3.1.7}
\end{equation}
Notice the non-central character of the extension. Had we worked with a central extension of the linear acceleration group, we would have got an exact homomorphism in \eqref{3.1.6}, with 
 zero for \eqref{3.1.7}. Here, even though the two-cocycle \eqref{2.9} leads to a non-central group extension, the cocycle representation of the linear acceleration group defined by \eqref{3.1.2} is \emph{not} 
 equivalent to a true representation of the extended group. In fact, it is this property that we will exploit to construct an associative, cocycle representation of the full Galilean line loop. 
 
It follows from \eqref{3.1.2} that the rest state vector $\mid\bs{0}\rangle$ is invariant under the action of rotations $\hat{U}^\times(0,R^0,\bs{0},0)$. Since the two-cocycle  \eqref{3.1.7} implies the factorization 
$\hat{U}^\times(0,R^0,\bs{a}_{\bs{q}},0)=\hat{U}^\times(0,I,\bs{a}_{\bs{q}},0)\hat{U}^\times(0,R^0,0,0)$, it then follows that $\hat{U}^\times(0,R^0,\bs{a}_{\bs{q}},0)$ and $\hat{U}^\times(0,I,\bs{a}_{\bs{q}},0)$
have the same action on the rest state vector $\mid\bs{0}\rangle$. In other words, as an element of the linear acceleration group, $g_{\bs{q}}$ is defined only up an arbitrary rotation and, just as the case for the Galilei group,  
the little group of a ${\cal G}$-orbit is $SO(3)$.  For further details of the construction of these representations, see \cite{bk1,sw1}. 

\subsection{Representations of the Galilean line loop}\label{sec3.2}
When rotations are time dependent, the set of constant space translations and prolongation parameters $\{(\varphi^0, I,\bs{a}^0,0)\}$ is not an invariant subgroup of $(\bar{\cal G}, \chi)$. Therefore, 
we cannot use the method of induced representations the same way as it was done for the linear acceleration group, starting with the constant spatial translation subgroup. As seen from \eqref{2.3.12}, the group of \emph{all} spatial 
translations  and prolongation parameters is a normal subgroup, but this group is not Abelian: 
\begin{eqnarray}
(\varphi_2,I,\bs{a}_2,0)(\varphi_1,I,\bs{a}_1,0)&=&\left(\varphi_2+\varphi_1+\frac{1}{2}(\bs{a}_2\cdot\dot{\bs{a}}_1-\dot{\bs{a}}_2\cdot\bs{a}_1),I,\bs{a}_2+\bs{a}_1,0\right)\nonumber\\
&\not=&\left(\varphi_2+\varphi_1-\frac{1}{2}(\bs{a}_2\cdot\dot{\bs{a}}_1-\dot{\bs{a}}_2\cdot\bs{a}_1),I,\bs{a}_2+\bs{a}_1,0\right)\nonumber\\
&=&(\varphi_1,I,\bs{a}_1,0)(\varphi_2,I,\bs{a}_2,0)\label{3.2.1}
\end{eqnarray}

Therefore, there are no common generalized eigenvectors of this invariant subgroup that we may use as a basis for the representation space. However, since the non-commutativity of \eqref{3.2.1} arises  
only in the loop prolongation parameter and 
since we only expect a representation where the homomorphism property holds  up to a phase --as seen from \eqref{3.1.6} and \eqref{3.1.7}, we do not get a true representation even when rotations are constant-- 
we may continue to treat $\mid\bs{q}\rangle$ as generalized eigenvectors of space translations, allowing homomorphism to hold up to a phase.  Furthermore, we make the reasonable demand that when rotations 
are suppressed, any cocycle representation of the Galilean line loop reduce to a representation of the extended linear acceleration group defined by \eqref{3.1.2}. Therefore, we only need to determine the action 
of rotations on the vectors $\mid\bs{q}\rangle$ because then the action of $\hat{U}^\times(\varphi,R,\bs{a},b)$ on $\mid\bs{q}\rangle$ follows from \eqref{3.1.2}: 
\begin{eqnarray}
\hat{U}^\times(\varphi,R,\bs{a},b)\mid\bs{q}\rangle&=&\hat{U}^\times\left((\varphi,I,\bs{a},b)(0,R,\bs{0},0)\right)\mid\bs{q}\rangle\nonumber\\
&=&\hat{U}^\times(\varphi,I,\bs{a},b)\hat{U}^\times(0,R,\bs{0},0)\mid\bs{q}\rangle\label{3.2.2}
\end{eqnarray} 
In writing the second equality of \eqref{3.2.2}, we have put an implicit constraint on the cocycle structure of the resulting representation, namely that the two-cochain $\xi_2(\bar{g}_2,\bar{g}_1,\bs{q})$ vanish when 
$\bar{g}_1$ is the of the form $(0,R,\bs{0},0)$.  This constraint is suggested by and consistent with the structure of the loop prolongation $(\bar{\cal G},\chi)$ because the two-cochain \eqref{2.9b} that defines 
it vanishes when either $g_2$ or $g_1$ has zero spatial translations. With such a two-cochain, we also have
\begin{equation}
\hat{U}^\times(\varphi,R,\bs{a},b)\mid\bs{q}\rangle=\hat{U}^\times(0,\Lambda_{-b}R,\bs{0},0)\hat{U}^\times(\varphi,I,R^{-1}\bs{a},b)\mid\bs{q}\rangle\label{3.2.3}
\end{equation}

The obvious choice for the action of rotations on $\mid\bs{q}\rangle$  is $\hat{U}^\times(0,R,\bs{0},0)|\bs{q}\rangle=|R\bs{q}\rangle$. Although the mapping 
$\bs{q}\to R\bs{q}$ is a homomorphism under rotations, when combined with the action of spatial translations, $\bs{q}\to\bs{q}+\dot{\bs{a}}$, it fails to be a homomorphism of the full Euclidean line group. 
The mapping that does lead to a homomorphism of the full Euclidean line group is
\begin{equation}
R:\ \bs{q}\to R\bs{q}+\dot{R}\bs{a}_{\bs{q}}=\frac{d}{dt}(R\bs{a}_{\bs{q}}),\label{3.2.4}
\end{equation}
where, as in \eqref{3.1.3}, $\bs{a}_{\bs{q}}=\int dt\,\bs{q}$.  Hence, let us define the action of time dependent rotations by 
\begin{equation}
\hat{U}^\times(0,R,\bs{0},0)\mid\bs{q}\rangle=\mid R\bs{q}+\dot{R}\bs{a}_{\bs{q}}\rangle\label{3.2.5}
\end{equation}
It then follows from \eqref{3.1.2}, \eqref{3.2.2} and \eqref{3.2.5},
\begin{equation}
\hat{U}^\times(\bar{g})\mid\bs{q}\rangle=e^{i\xi(\bar{g};\bs{q})}\mid\Lambda_{-b}\bs{q}'\rangle\label{3.2.6}
\end{equation}
where 
\begin{eqnarray}
\xi(\bar{g};\bs{q})&=&m\varphi+m\left(\bs{q}'\cdot\bs{a}-\frac{1}{2}\bs{a}\cdot\dot{\bs{a}}+\frac{1}{2}(\Lambda_{-b}-1)\bs{q}'\cdot\bs{a}_{\bs{q}'}\right)-wb\nonumber\\
\bs{a}_{\bs{q}}&=&\int dt\,\bs{q},\quad \bs{q}=\frac{d}{dt}\bs{a}_{\bs{q}}\nonumber\\
\bs{q}'&=&R\bs{q}+\dot{R}\bs{a}_{\bs{q}}+\dot{\bs{a}}=\frac{d}{dt}(R\bs{a}_{\bs{q}}+\bs{a})=\frac{d}{dt}\bs{a}_{\bs{q'}}\nonumber\\
w&=&\text{invariant, internal energy}\label{3.2.7}
\end{eqnarray}

In evaluating the integral $\int dt\,\bs{q}$ to determine the standard boost $\bs{a}_{\bs{q}}$, we choose the boundary condition that the arbitrary constant of integration is to be set to zero. For $\bs{q}$ that are 
analytic functions of time, $\bs{q}=\sum_{n=0}^\infty \frac{1}{n!}\bs{q}^{(n)}t^n$, this means choosing the boundary condition $\bs{a}_{\bs{q}}(t=0)=0$. 

Finally, by appealing to the duality formula \eqref{3.1.4}, we can deduce the unitary operators that furnish the representation on the set of square integrable wave functions:
\begin{equation}
\left(\hat{U}(\bar{g})\psi\right)(\bs{q})=e^{-i\xi(\bar{g}^{-1};\bs{q})}\psi(\Lambda_b\tilde{\bs{q}})\label{3.2.8}
\end{equation}
where 
\begin{equation}
\tilde{\bs{q}}=(\Lambda_{-b}R^{-1})\bs{q}+\Lambda_{-b}\dot{R}^{-1}\bs{a}_{\bs{q}}-\Lambda_{-b}(R^{-1}\dot{\bs{a}})-\Lambda_{-b}(\dot{R^{-1}}\bs{a})\label{3.2.9}
\end{equation}

\subsection{The associativity and cocycle structure of the representation}\label{sec3.3} 
The repeated application of \eqref{3.2.6} gives
\begin{equation}
\hat{U}^\times(\bar{g}_2)\hat{U}^\times(\bar{g}_1)\mid\bs{q}\rangle=e^{i\xi_2(\bar{g}_2,\bar{g}_1;\bs{q})}\hat{U}^\times(\bar{g}_2\bar{g}_1)\mid\bs{q}\rangle\label{3.3.1}
\end{equation}
where, with $\bs{q}'$ defined by \eqref{3.2.7} but with $\bar{g}=\bar{g}_1$,  
\begin{eqnarray}
\xi_2(\bar{g}_2,\bar{g}_1;\bs{q})&=&\xi(\bar{g}_1;\bs{q})+\xi(\bar{g}_2;\Lambda_{-b_1}\bs{q}')-\xi(\bar{g}_2\bar{g}_1;\bs{q})\nonumber\\
&=&m(1-\Lambda_{b_1})\varphi_2+m(\Lambda_{-b_1}-1)\omega_2(g_2,g_1g_{\bs{q}})\nonumber\\
&&\ +m(\Lambda_{b_1}\Omega_2)\cdot(R_1\bs{a}_{\bs{q}}\times\bs{a}_1-(\Lambda_{b_1}R_2)\bs{a}_1\times\Lambda_{b_1}\bs{a}_2)\label{3.3.2}
\end{eqnarray}
and $\omega_2$ is the two-cochain \eqref{2.9b}. 
When restricted to constant rotations, the angular velocity $\Omega=0$ and, as expected, \eqref{3.3.2} reduces to \eqref{3.1.7}. For constant rotations, constant $\varphi$, and spatial translations of the form $\bs{a}(t)=\bs{a}_0+\bs{v}t$, 
\eqref{3.3.2} vanishes, indicating that the representation reduces to a true representation of the centrally extended Galilei group. \emph{Furthermore, these reductions lead to the interpretation 
that the parameter $m$, the eigenvalue of the generator of the one parameter subgroup $\hat{U}^\times (\varphi, I,\bs{0},0)$, is the inertial mass.}\\

\noindent{\bf Associativity.} 
Let us now turn to the question of how to reconcile the violation of associativity of $(\bar{\cal G},\chi)$ and its obvious 
inevitability in the linear operator algebra to which the $\hat{U}^\times(\bar{g})$ belong. To that end, let us directly calculate the 
action of triple products $\hat{U}^\times(\bar{g}_3)\left(\hat{U}^\times(\bar{g}_2)\hat{U}^\times(\bar{g}_1)\right)$ and $\left(\hat{U}^\times(\bar{g}_3)\hat{U}^\times(\bar{g}_2)\right)\hat{U}^\times(\bar{g}_1)$ on an arbitrary vector 
$\mid\bs{q}\rangle$:
\begin{eqnarray}
\hat{U}^\times(\bar{g}_3)\left(\hat{U}^\times(\bar{g}_2)\hat{U}^\times(\bar{g}_1)\right)\mid\bs{q}\rangle&=&e^{i\xi_2(\bar{g}_2,\bar{g}_1;\bs{q})}\hat{U}^\times(\bar{g}_3)\hat{U}^\times(\bar{g}_2\bar{g}_1)\mid\bs{q}\rangle\nonumber\\
&=&e^{i\xi_2(\bar{g}_2,\bar{g}_1;\bs{q})+i\xi_2(\bar{g}_3,\bar{g}_2\bar{g}_1;\bs{q})}\hat{U}^\times\left(\bar{g}_3\left(\bar{g}_2\bar{g}_1\right)\right)\mid\bs{q}\rangle\nonumber\\ 
&=&e^{i\xi_2(\bar{g}_2,\bar{g}_1;\bs{q})+i\xi_2(\bar{g}_3,\bar{g}_2\bar{g}_1;\bs{q})}\hat{U}^\times\left(A(\bar{g}_3,\bar{g}_2,\bar{g}_1)\left[\left(\bar{g}_3\bar{g}_2\right)\bar{g}_1\right]\right)\mid\bs{q}\rangle\nonumber\\
&=&e^{i\xi_2(\bar{g}_2,\bar{g}_1;\bs{q})+i\xi_2(\bar{g}_3,\bar{g}_2\bar{g}_1;\bs{q})-i\xi_2(A(\bar{g}_3,\bar{g}_2,\bar{g}_1),(\bar{g}_3\bar{g}_2)\bar{g}_1;\bs{q})}\nonumber\\
&&\quad\times\hat{U}^\times(A(\bar{g}_3,\bar{g}_2,\bar{g}_1))\hat{U}^\times\left(\left(\bar{g}_3\bar{g}_2\right)\bar{g}_1\right)\mid\bs{q}\rangle\nonumber\\
\label{3.3.3}
\end{eqnarray}
where we have used \eqref{3.3.1} in the first, second and fourth steps and \eqref{2.3.6} in the third step. On the other hand, 
\begin{eqnarray}
\left(\hat{U}^\times(\bar{g}_3)\hat{U}^\times(\bar{g}_2)\right)\hat{U}^\times(\bar{g}_1)\mid\bs{q}\rangle&=&e^{i\xi(\bar{g}_1;\bs{q})}
\hat{U}^\times(\bar{g}_3)\hat{U}^\times(\bar{g}_2)\mid\Lambda_{-b_1}\bs{q}'\rangle\nonumber\\
&=&e^{i\xi_2(\bar{g}_3,\bar{g}_2;\Lambda_{-b_1}\bs{q}')}\hat{U}^\times(\bar{g}_3\bar{g}_2)\hat{U}^\times(\bar{g}_1)\mid\bs{q}\rangle\nonumber\\
&=&e^{i\xi_2(\bar{g}_3,\bar{g}_2;\Lambda_{-b_1}\bs{q}')+i\xi_2(\bar{g}_3\bar{g}_2,\bar{g}_1;\bs{q})}\hat{U}^\times\left((\bar{g}_3\bar{g}_2)\bar{g}_1\right)\mid\bs{q}\rangle\nonumber\\
&\equiv&e^{i\xi_2(\bar{g}_3,\bar{g}_2;g_1\bs{q})+i\xi_2(\bar{g}_3\bar{g}_2,\bar{g}_1;\bs{q})}\hat{U}^\times((\bar{g}_3\bar{g}_2)\bar{g}_1)\mid\bs{q}\rangle\nonumber\\
\label{3.3.4}
\end{eqnarray}
where we have used \eqref{3.2.6} in the first and second steps and \eqref{3.3.1} in the second and third steps. In the last step, we have merely introduced the notation $g\bs{q}=\Lambda_{-b}\bs{q}'$ that will be useful in the remainder of the calculation. 

Comparing \eqref{3.3.3} and \eqref{3.3.4}, we obtain
\begin{eqnarray}
\hat{U}^\times(\bar{g}_3)\left(\hat{U}^\times(\bar{g}_2)\hat{U}^\times(\bar{g}_1)\right)\mid\bs{q}\rangle&=&e^{i\xi_3(\bar{g}_3,\bar{g}_2,\bar{g}_1;\bs{q})}e^{-i\xi_2(A(\bar{g}_3,\bar{g}_2,\bar{g}_1), (\bar{g}_3\bar{g}_2)\bar{g}_1;\bs{q})}
\hat{U}^\times(A(\bar{g}_3,\bar{g}_2,\bar{g}_1))\nonumber\\
&&\quad\times\left[ \left(\hat{U}^\times(\bar{g}_3)\hat{U}^\times(\bar{g}_2)\right)\hat{U}^\times(\bar{g}_1)\right]\mid\bs{q}\rangle\label{3.3.5}
\end{eqnarray}
where 
\begin{equation}
\xi_3(\bar{g}_3,\bar{g}_2,\bar{g}_1;\bs{q})=\xi_2(\bar{g}_2,\bar{g}_1;\bs{q})+\xi_2(\bar{g}_3,\bar{g}_2\bar{g}_1;\bs{q})-\xi_2(\bar{g}_3,\bar{g}_2;g_1\bs{q})-\xi_2(\bar{g}_3\bar{g}_2,\bar{g}_1;\bs{q})\label{3.3.6}
\end{equation}
The square brackets in the last line of \eqref{3.3.5} may be omitted by the third axiom of Definition \ref{def2.2.2}.  We see that the three-cochain $\xi_3$ has the form of a coboundary. Since, as evident from from \eqref{3.3.2}, $\xi_2$ is itself a two-coboundary, $\xi_3$ would automatically vanish \emph{if} we were dealing with 
a representation of a group. However, for a loop, this is not necessarily the case, as may be seen by substituting (the first equality of) \eqref{3.3.2} in \eqref{3.3.6}:
\begin{equation}
\xi_3(\bar{g}_3,\bar{g}_2,\bar{g}_1;\bs{q})=\xi((\bar{g}_3\bar{g})\bar{g}_1;\bs{q})-\xi(\bar{g}_3(\bar{g}_2\bar{g}_1);\bs{q})\label{3.3.6b}
\end{equation}
The non-associativity of $(\bar{\cal G},\chi)$ means that \eqref{3.3.6b} does not necessarily vanish. Using \eqref{3.3.2} again in the other exponent of \eqref{3.3.5}, 
\begin{eqnarray}
\xi_2(A(\bar{g}_3,\bar{g}_2,\bar{g}_1), (\bar{g}_3\bar{g}_2)\bar{g}_1;\bs{q})&=&\xi((\bar{g}_3\bar{g}_2)\bar{g}_1;\bs{q})+\xi(A(\bar{g}_3,\bar{g}_2,\bar{g}_1);g_3g_2g_1\bs{q})\nonumber\\
&&\quad-\xi(A(\bar{g}_3,\bar{g}_2,\bar{g}_1)[(\bar{g}_3\bar{g}_2)\bar{g}_1];\bs{q})\nonumber\\
&=&\xi((\bar{g}_3\bar{g}_2)\bar{g}_1;\bs{q})-\xi(\bar{g}_3(\bar{g}_2\bar{g}_1);\bs{q})\nonumber\\
&&\quad+\xi(A(\bar{g}_3,\bar{g}_2,\bar{g}_1);g_3g_2g_1\bs{q})\label{3.3.7}
\end{eqnarray}
Therefore, substituting \eqref{3.3.6b} and \eqref{3.3.7} in \eqref{3.3.5}, we obtain the identity
\begin{eqnarray}
\hat{U}^\times(\bar{g}_3)\left(\hat{U}^\times(\bar{g}_2)\hat{U}^\times(\bar{g}_1)\right)\mid\bs{q}\rangle&=&e^{-i\xi(A(\bar{g}_3,\bar{g}_2,\bar{g}_1);g_3g_2g_1\bs{q})}
\hat{U}^\times(A(\bar{g}_3,\bar{g}_2,\bar{g}_1))\nonumber\\
&&\quad\left(\hat{U}^\times(\bar{g}_3)\hat{U}^\times(\bar{g}_2)\right)\hat{U}^\times(\bar{g}_1)\mid\bs{q}\rangle\label{3.3.8}
\end{eqnarray}

So far, our analysis of associativity of the representation has been quite general. In particular, we have not used the specific form of  \eqref{3.2.7} that defines the representation anywhere in \eqref{3.3.3}-\eqref{3.3.8}. 
In fact, since the operators $U^\times(\bar{g})$ necessarily associate, we may view \eqref{3.3.8}  a constraint, 
\begin{eqnarray}
\hat{U}^\times(A(\bar{g}_3,\bar{g}_2,\bar{g}_1))\hat{U}^\times(\bar{g}_3)\hat{U}^\times(\bar{g}_2)\hat{U}^\times(\bar{g}_1)\mid\bs{q}\rangle&=&
e^{i\xi(A(\bar{g}_3,\bar{g}_2,\bar{g}_1);g_3g_2g_1\bs{q})}\nonumber\\
&&\ \hat{U}^\times(\bar{g}_3)\hat{U}^\times(\bar{g}_2)\hat{U}^\times(\bar{g}_1)\mid\bs{q}\rangle,\nonumber\\
\label{3.3.9}
\end{eqnarray}
on any one-cocahin $\xi(\bar{g};\bs{q})$ that defines the representation. 

For our particular representation, it follows from \eqref{3.2.6} and \eqref{3.2.7} that for any loop element in the kernel of $\chi$, i.e., $\bar{g}=(\varphi,e)$, 
the operator $\hat{U}^\times(\varphi,e)$ is simply multiplication by the phase $e^{im\varphi}$. In particular, for $\bar{g}=(\varphi,e)$, \eqref{3.2.7} reduces to 
\begin{equation}
\xi((\varphi,e);\bs{q})=m\varphi\label{3.3.10}
\end{equation}
Since all associators lie in the kernel of $\chi$, with \eqref{2.3.7}, we then have
\begin{eqnarray}
\xi(A(\bar{g}_3,\bar{g}_2,\bar{g}_1;\bs{q}))&=&m\Lambda_{-(b_3+b_2+b_1)}(\delta_2\omega_2)(g_3,g_2,g_1)\label{3.3.11}\\
\hat{U}^\times(A(\bar{g}_3,\bar{g}_2,\bar{g}_1))\mid\bs{q}\rangle&=&e^{im\Lambda_{-(b_3+b_2+b_1)}(\delta_2\omega_2)(g_3,g_2,g_1)}\mid\bs{q}\rangle\label{3.3.12}
\end{eqnarray}
Therewith, \eqref{3.3.5} becomes
\begin{equation}
\hat{U}^\times(\bar{g}_3)\left(\hat{U}^\times(\bar{g}_2)\hat{U}^\times(\bar{g}_1)\right)\mid\bs{q}\rangle
= \left(\hat{U}^\times(\bar{g}_3)\hat{U}^\times(\bar{g}_2)\right)\hat{U}^\times(\bar{g}_1)\mid\bs{q}\rangle.\label{3.3.13}
\end{equation}
This shows that what we have constructed is an associative, cocycle representation of the non-associative Galilean line loop. In particular, all associators are represented 
by ($\bs{q}$-independent) phases that are precisely cancelled by the three-cobundary of the corresponding triple product.

\subsection{The Hamiltonian}\label{sec3.4}
Although $(\bar{\cal G},\chi)$ is a loop, the set of time translations, space translations and rotations are subgroups of $(\bar{\cal G},\chi)$. Therefore, we may define the physical observables of Hamiltonian, momenta and angular momenta, respectively, as generators of these subgroups the usual way.  In particular, the Hamiltonian is
\begin{equation}
\hat{H}:=\left.i\frac{\hat{U}(0,I,\bs{0},b)}{b}\right|_{b=0}.\label{3.4.1}
\end{equation}
We may obtain the realization of $\hat{H}$ as an operator in the space of wave functions from \eqref{3.4.1} and the transformation  formula \eqref{3.2.8}:
\begin{eqnarray}
\left(\hat{H}\psi\right)(\bs{q})&=&i\left.\frac{d}{db}\left(\hat{U}(0,I,\bs{0},b)\psi\right)(\bs{q})\right|_{b=0}\nonumber\\
&=&i\left.\frac{d}{db} e^{-i\frac{1}{2}m(\Lambda_b-1)\bs{q}\cdot\bs{a}_{\bs{q}}-iwb}\psi(\Lambda_b\bs{q})\right|_{b=0}\nonumber\\
&=&\left(\frac{1}{2}m\bs{q}^2+w+\frac{1}{2}m\dot{\bs{q}}\cdot\bs{a}_{\bs{q}}+i\dot{\bs{q}}\cdot\nabla_{\bs{q}}\right)\psi(\bs{q})\label{3.4.2}
\end{eqnarray}

Similarly, we define momenta as the generators of time independent spatial translations $\bs{a}^{0}$: 
\begin{equation}
\hat{P}_i=-\left.i\frac{\hat{U}(0,I,a^{0}_i)}{a^{0}_i}\right|_{a^0_i=0}.\label{3.4.3}
\end{equation}
Again, we obtain the realization of $\hat{\bs{P}}$ as operators in the space of wave functions from \eqref{3.2.8}:
\begin{equation}
\left(\hat{\bs{P}}\psi\right)(\bs{q})=m\bs{q}\psi(\bs{q}).\label{3.4.4}
\end{equation}
This expression justifies our interpretation of $\bs{q}$ as velocity. It also suggests that we continue to define  position operators, 
\begin{equation}
\hat{\bs{X}}=i\nabla_{\bs{p}}=i\frac{1}{m}\nabla_{\bs{q}}\label{3.4.5}
\end{equation}
so that they are canonically conjugated to the $\hat{\bs{P}}$. Obviously, these are not the only operators that are canonically conjugated to the momentum operators and can serve as position operators. 
For a discussion on this matter, see \cite{bk1}. 

In terms of the position and momentum operators, we may express the Hamiltonian as follows: 
\begin{equation}
\left(\hat{H}\psi\right)(\bs{q})=\left(\frac{\hat{\bs{P}}^2}{2m}+\hat{V}+m\dot{\bs{q}}\cdot\left(\hat{\bs{X}}+\frac{1}{2}\bs{a}_{\bs{q}}\right)\right)\psi(\bs{q})\label{3.4.6}
\end{equation}
where we have introduced $\hat{V}$ as the internal energy operator, an invariant by construction. Its eigenvalues span the range of the internal energy parameter $w$. 

The implications of \eqref{3.4.6} for the role of the equivalence principle in quantum mechanics were discussed in \cite{sw1}. Further developments showing the violations 
of the equivalence principle were reported in \cite{bksw}. Therefore, here we will consider only the effects of rotationally accelerating reference frames on the Hamiltonian. 

Consider a transformation from an inertial reference frame to a rotating reference frame. In an inertial frame, the velocity variables $\bs{q}^0$ are time independent. Therefore, the standard boost 
that maps from zero velocity to $\bs{q}^0$ is of the form $\bs{a}_{\bs{q}^{0}}=\bs{q}^{0}t$.  Then, from \eqref{3.2.5}, 
\begin{equation}
\mid\bs{q}\rangle=\hat{U}^\times(0,R,\bs{0},0)\mid\bs{q}^{0}\rangle\label{3.4.7}
\end{equation}
where 
\begin{eqnarray}
\bs{q}&=&R\bs{q}^{0}+\dot{R}\bs{a}_{\bs{q}^{0}}\nonumber\\
&=&\frac{d}{dt}(R\bs{a}_{\bs{q}^0})\nonumber\\
&=&R\bs{q}^0+\dot{R}R^T\bs{a}_{\bs{q}}\nonumber\\
&=&R\bs{q}^{0}+\Omega\times\bs{a}_{\bs{q}}\label{3.4.8}
\end{eqnarray}
Differentiating this expression,
\begin{eqnarray}
\dot{\bs{q}}&=&\dot{R}\bs{q}^{0}+\dot{\Omega}\times\bs{a}_{\bs{q}}+\Omega\times\bs{q}\nonumber\\
&=&\dot{R}R^T(\bs{q}-\Omega\times\bs{a}_{\bs{q}})+\dot{\Omega}\times\bs{a}_{\bs{q}}+\Omega\times\bs{q}\nonumber\\
&=&\dot{\Omega}\times\bs{a}_{\bs{q}}+2\Omega\times\bs{q}-\Omega\times(\Omega\times\bs{a}_{\bs{q}})\label{3.4.9}
\end{eqnarray}

We can obtain the contributions to the Hamiltonian due to the non-inertial nature of the rotating reference frame by substituting \eqref{3.4.9} in \eqref{3.4.6}. 
It is clear from \eqref{3.4.9} that we recover both the centrifugal force term $-m\Omega\times(\Omega\times\bs{a}_{\bs{q}})$ and the Coriolis force term $2m\Omega\times\bs{q}$. 
In the Hamiltonian, both of these forces are multiplied by the position operator, leading to fictitious potential energy terms. (The appearance of  the term $\frac{1}{2}\bs{a}_{\bs{q}}$ that gets 
added to the position operator is not significant. It is a the result of our choice of the two-cochain \eqref{2.9b} and may be removed by adding a coboundary to \eqref{2.9b}.) 
As expected, these fictitious potential energy terms are proportional to the inertial mass.

\section{Concluding remarks}\label{sec4}
In this paper we have presented an analysis of the Galilean line group, the group of transformations amongst all rotationally and linearly accelerating 
reference frames in a Galilean spacetime, and constructed its general unitary cocycle representations. As shown in Section \ref{sec3}, these representations 
are defined by the transformation formula for the generalized velocity eigenvectors $\mid\bs{q}\rangle$, 
\begin{equation}
\hat{U}^\times(\bar{g})\mid\bs{q}\rangle=e^{i\xi(\bar{g};\bs{q})}\mid\Lambda_{-b}\bs{q}'\rangle\label{4.1}
\end{equation}
where 
\begin{eqnarray}
\bar{g}&=&(\varphi,R,\bs{a},b)\nonumber\\
\xi(\bar{g};\bs{q})&=&m\varphi+m\left(\bs{q}'\cdot\bs{a}-\frac{1}{2}\bs{a}\cdot\dot{\bs{a}}+\frac{1}{2}(\Lambda_{-b}-1)\bs{q}'\cdot\bs{a}_{\bs{q}'}\right)-wb\nonumber\\
\bs{a}_{\bs{q}}&=&\int dt\,\bs{q},\quad \bs{q}=\frac{d}{dt}\bs{a}_{\bs{q}}\nonumber\\
\bs{q}'&=&R\bs{q}+\dot{R}\bs{a}_{\bs{q}}+\dot{\bs{a}}=\frac{d}{dt}(R\bs{a}_{\bs{q}}+\bs{a})=\frac{d}{dt}\bs{a}_{\bs{q'}}\nonumber\\
w&=&\text{invariant, internal energy}\label{4.2}
\end{eqnarray}
Here, $R$ and $\bs{a}$ are, respectively, time dependent rotations and space translations. They furnish transformations into arbitrarily accelerating reference frames. 

The main physical constraint under which the representations \eqref{4.1} were constructed was that each  reduces to a unitary \emph{projective} representation 
of the Galilei group  when the acceleration transformations are restricted to the subgroup of inertial (i.e., Galilei) transformations. This condition imposes 
an intricate cocycle structure on  \eqref{4.1}. As seen from \eqref{3.3.1} and \eqref{3.3.2}, the representation \eqref{4.1} is a two-cocycle representation where 
the phase factor that relates the action of operators $\hat{U}^\times(\bar{g}_2\bar{g}_1)$ and $\hat{U}^\times(\bar{g}_2)\hat{U}^\times(\bar{g}_1)$ on a vector $\mid\bs{q}\rangle$ 
depends not only on the group parameters but also on the value of $\bs{q}$. This situation is in sharp contrast to the projective representations of the Galilei group where the 
phase factor is given by the $\bs{q}$-independent function \eqref{2.12}. 

The form of the above phase factor has important implications for the construction of the representations. In particular, \eqref{2.12} and the cocycle condition \eqref{2.11} that it fulfills 
make a physical representation of the Galilei group equivalent to a vector representation of a central extension of the Galilei group. Furthermore, this central extension
 contains an Abelian invariant subgroup. Therefore, we can use the method of induced representations to obtain the representations of the central extension,  and therewith 
the relevant projective representations of the Galilei group.  This is not the case for the Galilean line group. As shown in Section \ref{sec2}, in particular \eqref{2.13}, there are no 
functions $\omega:\ {\cal G}\times{\cal G}\to{\cal F}(\mathbb{R})$ that fulfill the dual requirement of the two-coycle condition \eqref{2.11} and the reduction to the Galilei 
two-cocycle \eqref{2.12}. The absence of two-cocycles of the Galilean line group means that \emph{there exist no group extensions whose true representations are equivalent to 
the general cocycle representations \eqref{4.1}}. Such group extensions are possible for the linear acceleration group ${\cal A}_{linear}$ (i.e., without time dependent rotations), 
albeit these extensions are non-central.  Even here, owing to the non-centrality of the extension, the situation is different from the Galilei group: as seen 
from \eqref{3.1.6} and \eqref{3.1.7}, the two-cocycle representations of the linear acceleration group ${\cal A}_{linear}$ are not equivalent to true representations of its 
non-central extensions. 

While there exist no group extensions of the Galilean line group by ${\cal F}(\mathbb{R})$, either central or non-central, 
we have shown that there exist \emph{loop prolongations} of the Galilean line group by ${\cal F}(\mathbb{R})$. The representations \eqref{4.1} were obtained as 
representations of these loop prolongations. Even as a representation of the loop, as \eqref{3.3.1} and \eqref{3.3.2} show, \eqref{4.1} furnishes a cocycle representation, not a true homomorphism 
$\bar{g}\to \hat{U}^\times(\bar{g})$.  The simple conclusion 
to be drawn from these observations  is that the equivalence of the two-cocycle representations to true representations of the central extensions is a result that holds only in the special case of 
the Galilean transformations. Whenever acceleration transformations are brought into the picture, the situation becomes much more complicated. 

We obtained the representations \eqref{4.1} as cocycle representations of the loop prolongation of the Galilean line group constructed in Section \ref{sec2.2} by using the same 
well-known method of induced representations of groups.  A rather remarkable feature of the resulting representations is that, as seen from \eqref{3.3.13}, 
 they are completely consistent with the necessary property of associativity 
 of an algebra of linear 
operators, despite the violation of associativity for the Galilean line loop \eqref{2.3.6}. In fact, we identify \eqref{3.3.9} as a general constraint 
on any cochain representation of a loop by linear operators in a vector space. Therefore, we see our analysis as an extension of the method of induced representations to a class of loops, a set of 
objects more general than groups. Hence, the mathematical utility of the method may extend beyond the present study.  Our analysis also shows that we should 
allow for the possibility that quantum mechanically relevant symmetry transformations, be they spacetime transformations or internal symmetry transformations, are encoded by loops, rather than groups. 
As such, the construction presented here may have greater generality and value beyond the its use for formulating quantum mechanics in non-inertial reference frames. 

As final remark, we note that we have made a specific choice of two-cochain \eqref{2.9b} for our construction of the Galilean line loop. This cochain is not a two-cocycle, but it is a three-cocycle. 
The third cohomology group of the Galilean line group is in fact infinite dimensional, containing elements of the form \eqref{2.14}. There exist loop prolongations of the Galilean 
line group, of the same form as, but inequivalent to, those constructed here,  corresponding to distinct elements of the third cohomology group. Unitary representations of these loops 
can be constructed the same way as done here and the violations of the equivalence principle shown in \cite{bksw} can also be seen from these general loop representations. That the equivalence 
principle is completely upheld for the representations developed here is a special attribute of the cochain~\eqref{2.9b}.

\appendix
\section{Appendix: Cohomology, group extensions and loop prolongations}\label{appendix}
Ever since the analysis of Bargmann, it is known that a projective representation of the Galilei group is equivalent to 
a true (vector) representation of a suitable central extension of the group. In this sense, it is the central extensions of the Galilei group, rather than the group itself, that is of importance 
in quantum mechanics. As shown in Section \ref{sec2}, it is therefore necessary that we construct suitable extensions of the Galilean line group in which 
a central extension of the Galilei group that corresponds to a given projective representation may be embedded. In this Appendix we collect some results that provide the 
mathematical framework for our analysis of Section \ref{sec2}. Our focus is on the algebraic aspects of cohomology of groups and loops and we will not 
concern ourselves with the topological issues. 

\subsection{Group extensions}\label{Appexdix-sec3.1} 
 The general problem of extending a group $G$ by another group $A$ can be stated as follows: find a group $\tilde{G}$ such that $A$ is an invariant subgroup of $\tilde{G}$ and the quotient 
 $\frac{\tilde{G}}{A}$ is isomorphic to $G$. Hence, we may think of $\tilde{G}$ as the principal fiber bundle with structure group $A$. 
 Then, if we choose a normalized trivializing section $s: G\to\tilde{G}$, an element $\tilde{g}$ of 
 $\tilde{G}$ may be written as $\tilde{g}=as(g)\equiv (a,g)$ for some $a\in A$ and $g\in G$. The composition rule for $\tilde{G}$ reads
 \begin{eqnarray}
 \tilde{g}_2\cdot\tilde{g}_1&=&a_2s(g_2)a_1s(g_1)\nonumber\\
 &=&a_2(\sigma(g_2)a_1)\omega(g_2,g_1)s(g_2g_1)\nonumber\\
&\equiv&(a_2,g_2)(a_1,g_1)\nonumber\\
&=&(a_2(\sigma(g_2)a_1)\omega_2(g_2,g_1),g_2g_1)\label{3.1}
 \end{eqnarray}
where $\sigma(g)a=s(g)as(g)^{-1}$ is the automorphism of $A$ defined by $s(g)\in\tilde{G}$ and $\omega_2(g_2,g_1)$ is a unique element of $A$. 
The associativity of the product rule \eqref{3.1} under which $\tilde{G}$ is a group imposes the following 
restriction on  $\omega_2: \ G\times G\to A$: 
\begin{equation}
\omega_2(g_3,g_2)\omega_2(g_3g_2,g_1)=\left(\sigma(g_3)\omega_2(g_2,g_1)\right)\omega_2(g_3,g_2g_1)\label{3.2}
\end{equation}
A function $\omega_2: \ G\times G\to A$ satisfying \eqref{3.2} is called a \emph{factor system} relative to (normalized) trivializing section $s$.  What is of interest to us in this paper involves the case where 
$A$ is an Abelian group. Then, writing \eqref{3.2} additively, we have 
\begin{equation}
\omega_2(g_3,g_2)+\omega_2(g_3g_2,g_1)=\sigma(g_3)\omega_2(g_2,g_1)+\omega_2(g_3,g_2g_1)\label{3.3}
\end{equation}
and therewith the composition rule \eqref{3.1} for $\tilde{G}$ becomes, 
\begin{equation}
(a_2,g_2)(a_1,g_1)=(a_2+\sigma(g_2)a_1+\omega_2(g_2,g_1), g_2g_1)\label{3.3b}
\end{equation}
For a given automorphism $\sigma(g)$, all inequivalent extensions of  $G$ by $A$ are determined by inequivalent mappings $\omega_2:\ G\times G\to A$. (Note that the discrepancy in the placement of the automorphism between  \eqref{2.10} and \eqref{3.3b} and between \eqref{2.11} and \eqref{3.3} is due to the composition structure of ${\cal G}$. See footnote \ref{footnote1}.) 

\subsection{The second cohomology group}\label{Appendix-sec3.2}
The collection of inequivalent mappings that fulfill \eqref{3.3} constitute the second cohomology group $H_{\sigma}^2(G,A)$. In order to define it, we must first introduce the key notions of cochains, cocycles and coboundaries.Throughout this subsection, we let $A$ be an Abelian group and $G$, an arbitrary group.\\
{\bf n-cochains:} The mapping  $\alpha_n:\ G\times G\times\cdots\times G\to A$, i.e., 
\begin{equation}
\alpha_n: (g_1,g_2,\cdots, g_n)\to \alpha_n(g_1,g_2,\cdots, g_n)\in A\label{3.4}
\end{equation}
 is called an $n$-cochain. The collection of $n$-cochains form an Abelian group that we denote by $C^n(G,A)$, its composition rule defined by the pointwise composition of $\alpha_n(g_1,g_2,\cdots, g_n)$ as elements $A$. \\
{\bf Coboundary operator:} Let $\sigma:\ G\to \text{Aut}\ A$ be a homomorphism from $G$ to the automorphism group on $A$. 
The operator  $\delta_n:\ C^n(G,A)\to C^{n+1}(G,A)$  defined by the mapping (for the left action)
\begin{eqnarray}
(\delta_n\alpha_n)(g_1,g_2,\cdots,g_n,g_{n+1})&:=&\sigma(g_1)\alpha_n(g_2,g_3,\cdots,g_{n+1})\nonumber\\
&&\quad+\sum_{i=1}^n(-1)^i\alpha_n(g_1,\cdots,g_{i-1},g_i\cdot g_{i+1}, g_{i+2},\cdots, g_{n+1})\nonumber\\
&&\qquad +(-1)^{n+1}\alpha_n(g_1,g_2,\cdots,g_n)\label{3.5}
\end{eqnarray}
is called the coboundary operator. It is straightforward, albeit a bit tedious, to show that 
\begin{equation}
\delta_{n+1}\circ\delta_n=0\label{3.6}
\end{equation}
Hence, 
\begin{equation}
\text{range}\ \delta_n\subset\ \text{kernel}\ \delta_{n+1}\label{3.7}
\end{equation}
{\bf Cocycles and coboundaries:} The inclusion \eqref{3.7} motivates the following definitions:
\begin{eqnarray}
\text{$n$-cocycles:}&&Z_\sigma^n(G,A):=\text{kernel}\ \delta_n\nonumber\\
\text{$n$-coboundaries:}&& B_\sigma^n(G,A):=\text{range}\ \delta_{n-1}\label{3.8}
\end{eqnarray}
Both $Z_\sigma^n(G,A)$ and $B_\sigma^n(G,A)$ are subgroups of $C^n(G,A)$. Further, by \eqref{3.7}, every $n$-coboundary is trivially an $n$-cocycle and we so have the inclusion 
$B_\sigma^n(G,A)\subset Z_\sigma^n(G,A)$. Since $A$ is Abelian, $B_\sigma^n(G,A)$ is in fact  an invariant subgroup of $Z_\sigma^n(G,A)$.\\
{\bf The $\bs{n^{\rm th}}$ cohomology group:} The quotient group
\begin{equation}
H_\sigma^n(G,A)=\frac{Z^n_\sigma(G,A)}{B_\sigma^n(G,A)}\label{3.9}
\end{equation}
is called the $n^{\rm th}$-cohomology group. An element of $H_\sigma^n(G,A)$ is a class of $n$-cocycles, equivalent up to $n$-coboundaries.\\
{\bf The second cohomology group:} A comparison of  \eqref{3.3} with the general formula \eqref{3.5} and the cocycle condition \eqref{3.8} shows 
that the constraint on the factor system that defines an extension of $G$ by $A$ is precisely the two-cocycle condition:
\begin{equation}
(\delta_2\alpha_2)(g_3,g_2,g_1):=\sigma(g_3)\alpha_2(g_2,g_1)-\alpha_2(g_3g_2,g_1)+\alpha_2(g_3,g_2g_1)-\alpha_2(g_3,g_2)=0\label{3.10}
\end{equation}
Moreover, from \eqref{3.3b} and \eqref{3.10}, it is possible to show that two cocycles $\alpha_2$ and $\alpha'_2$ that differ by a two-coundary, 
\begin{eqnarray}
\alpha_2(g_2,g_1)-\alpha'_2(g_2,g_1)&=&(\delta_1\alpha_1)(g_2,g_1)\nonumber\\
&=&\sigma(g_2)\alpha_1(g_1)-\alpha_1(g_2g_1)+\alpha_1(g_2),\label{3.11}
\end{eqnarray}
lead to equivalent extensions of $G$ by $A$. This means that, for a given $\sigma:\ G\to\text{Aut}\ A$, inequivalent extensions of $G$ by $A$ are in one-to-one correspondence with 
the elements of the second cohomology group $H_\sigma^2(G,A)$.   Note that while $A$ is an invariant subgroup of $\tilde{G}$, in general, $G$ is \emph{not} a subgroup of 
$\tilde{G}$. 

There are three special cases commonly encountered in the physics literature: 
\begin{enumerate}
\item {\bf Trivial $\sigma$ and non-trivial $\omega_2\in H^2(G,A)$:} In this case, the group law \eqref{3.3} becomes
\begin{equation}
(a_2,g_2)(a_1,g_1)=(a_2+a_1+\omega_2(g_2,g_1), g_2g_1)\label{3.12}
\end{equation}
In this case,  $\tilde{G}$ is called a \emph{central extension} of $G$ by $A$ and $A$ is a central subgroup of $\tilde{G}$. In particular, 
an extension of $G$ by $\mathbb{R}$ is central. The necessity that the automorphism $\sigma(g)$ be trivial for the extension to be central highlights the 
obverse case considered in Section \ref{sec2}. It is the non-triviality of the automorphism $\sigma(R,\bs{a},b)=\Lambda_b$ that makes the extension  \eqref{2.10} 
of the linear acceleration group non-central. 
\item {\bf Non-trivial $\sigma$ and trivial $\omega_2=0\in H_\sigma^2(G,A)$:} In this case, \eqref{3.3} becomes 
\begin{equation}
(a_2,g_2)(a_1,g_1)=(a_2+\sigma(g_2)a_1, g_2g_1)\label{3.13}
\end{equation}
This is the semidirect product of $G$ by $A$. In this case, $G$ is also a subgroup of $\tilde{G}$. 
\item {\bf Trivial $\sigma$ and trivial $\omega_2$:} The group law \eqref{3.3} now reads
\begin{equation}
(a_2,g_2)(a_1,g_1)=(a_2+a_1,g_2g_1)\label{3.14}
\end{equation}
This is the direct product of $G$ and $A$, both being invariant subgroups of $\tilde{G}$. 
\end{enumerate}

\subsection{Loops and loop prolongations}
\noindent{\bf Properties of loops.} Let $L$ be a loop, fulfilling the properties of Definition \ref{def2.2.1}. 
A subset $S\subset L$ that is closed under the multiplication rule of $L$ is a subloop of $L$. If all associators, defined by \eqref{2.2.1}, reduce to the identity on $S$, then $S$ is a subgroup of $L$. 
Given a subgroup $S$, we may define left and right cosets $aS$ and $Sa$ as usual. We call a subgroup $S$ of $L$ \emph{normal} if $aS=Sa$ for every $a\in L$. 
For a normal subgroup of a group, the definition can be equivalently given in terms of the invariance under inner automorphisms $s\to asa^{-1}$. For a loop, as a result \eqref{2.2.1}, the 
triple product $asa^{-1}$ is not uniquely defined. However, from the third axiom of Definition \ref{def2.2.1}, for any $a,b\in L$, there exists an element 
$a\circ b\in L$ such that 
\begin{equation}
ab=(a\circ b)a\label{2.2.2}
\end{equation}
For every $a\in L$, the bijection $b\to a\circ b$ is called the \emph{inner transformation} induced by the element $a$. It then follows that a subgroup $S$ is normal if and only if every inner transformation 
maps $S$ onto itself.

If $S$ is a normal subgroup of a loop $L$ and if $A(s,a,b)=A(a,s,b)=1$ for all $s\in S$ and $a,b\in L$, then it is a straightforward to show 
that the set of right cosets of $S$ forms a loop $L/S$ under the multiplication rule  $(Sa)(Sb)=S(ab)$. Further, the mapping $a\to Sa$ is a homomorphism of $L$ onto $L/S$, the kernel 
of which is $S$. 

\noindent{\bf Loop prolongations.} Let $(L,\chi)$ be a loop prolongation of a group $G$ by an Abelian group $K$. The pair $(L,\chi)$ fulfills the properties of Definition \eqref{def2.2.2}, which we restate 
here as we will repeatedly refer to it. 
{\defin{Let $G$ be a group, $K$ an Abelian group and $\sigma(g)$ an automorphism on $K$ for every $g\in G$. Then, the pair $(L,\chi)$ is called a \emph{loop prolongation} 
of $G$ by $K$ if the following axioms hold:}
\begin{enumerate}
\item $L$ is a loop such that (the canonical injection of) $K$ is a subgroup of $L$. 
\item $\chi$ is a homomorphism of $L$ onto $G$ with kernel $S$ containing ${K}$. 
\item $A(s,a,b)=A(a,s,b)=1$ for all $s\in S$, $a,b,\in L$. 
\item $A(a,b,c)s=sA(a,b,c)$ for all $s\in S$,  $a,b,c \in L$. 
\item $a\circ{k}={\sigma(\chi(a)^{-1})k}$ for all $a\in L$ and $k\in K$. 
\end{enumerate}\label{def2.2.3}}
The third and fourth conditions, which put limits the non associativity of $L$, show that a triple product in which the first or second element belongs to 
$S$ is associative.  In particular, associativity holds for any triple product in which the first or second element is an associator.  

The following assertions are direct consequences of Definition \ref{def2.2.3}:
\begin{eqnarray}
&&1. \ \text{The kernel}\ S\ \text{of}\ \chi\ \text{is a normal subgroup of}\ L\label{2.2.3}\\
&&2.\ \text{The center}\ Z\ \text{of}\ S\ \text{is a normal subgroup of}\ L.\label{2.2.4}\\
&&3.\  K\subset Z\ \text{and}\ A(a,b,c)\in Z,\ \text{for all}\ a,b,c\in L\label{2.2.5}\\
&& 4.\ A(k,a,b)=A(a,k,b)=A(a,b,k)=1\label{2.2.6}
\end{eqnarray}
The last property, in particular, means that associativity holds in any triple product that involves an element of $K$. 

Associators fulfill an identity that mirrors the three-cocycle condition (i.e., $\alpha_3$ is annihilated by the coboundary operator defined by \eqref{3.5}). 
For $a,b,c,d\in L$, using the third axiom of Definition \ref{def2.2.3} to express 
$a[b(cd)]$ in terms of $[(ab)c]d$ and associators in two different ways, we obtain
\begin{equation}
a\circ A(b,c,d)-A(ab,c,d)+A(a,bc,d)-A(a,b,cd)+A(a,b,c)=0.\label{2.2.7}
\end{equation}
We have written \eqref{2.2.7} additively since all associators are in the center $Z$ of the kernel $S$. 

The associators in fact define a characteristic three coboundary of $G$ that takes values in $Z$. To see this, let $\tau:\ G\to L$ be a section, normalized so that $\tau(1)=1$. 
Then, by the third axiom of Definition \ref{def2.2.1}, there exists an element $\omega^\tau_2(g_2,g_1)$ such that 
\begin{equation}
\tau(g_1)\tau(g_2)=\omega_2^\tau(g_1,g_2)\tau(g_1g_2), \ g_1,g_2\in G\label{2.2.8}
\end{equation}
The application of the homomorphism $\chi$ to \eqref{2.2.8} shows that $\omega_2^\tau(g_1,g_2)\in S$. Hence, it is a \emph{factor system} corresponding to the section $\tau$. 
The normalization of $\tau$ implies that $\omega_2^\tau(g,1)=\omega^\tau_2(1,g)=1$.
A direct computation using \eqref{2.2.1} and \eqref{2.2.8} yields
\begin{equation}
A(\tau(g_1),\tau(g_2),\tau(g_3))\omega_2^\tau(g_1,g_2)\omega_2^\tau(g_1g_2,g_3)=[\tau(g_1)\circ\omega_2^\tau(g_2,g_3)]\omega_2^\tau(g_1,g_2g_3)\label{2.2.9}
\end{equation}
In the commutative case, this equality can be written as 
\begin{equation}
A(\tau(g_1),\tau(g_2),\tau(g_3))=[\tau(g_1)\circ\omega_2^\tau(g_2,g_3)]+\omega_2^\tau(g_1,g_2g_3)-\omega_2^\tau(g_1,g_2)-\omega_2^\tau(g_1g_2,g_3)\label{2.2.10}
\end{equation}
This expression should be compared with \eqref{2.11} and \eqref{2.13}. In fact, \eqref{2.2.10} motivates the following definition of a characteristic three-cochain on $G$ with values in  $Z$: 
\begin{equation}
\omega_\tau^3(g_1,g_2,g_3)=A(\tau(g_1), \tau(g_2),\tau(g_3))\label{2.2.11}
\end{equation}

Higher order associators can be defined by induction, and therewith also higher order characteristic cochains of $G$ with values in $Z$ corresponding to a section $\tau:\ G\to L$: 
\begin{eqnarray}
A(a_1,\cdots,a_{2n}, a)&:=&A(a_1,a_2,A(a_3,\cdots,a_{2n},a))\nonumber\\
\omega_\tau^{2n+1}(g_1,\cdots,g_{2n},g)&:=&A(\tau(g_1),\cdots,\tau(g_{2n}),\tau(g)),\quad n\geq1\nonumber\\
\omega_\tau^{2n+2}(g_1,\cdots,g_{2n},g, h)&:=&A(\tau(g_1),\cdots,\tau(g_{2n}),\omega_\tau^2(g,h)),\quad n\geq1\nonumber\\ \label{2.2.12}
\end{eqnarray}

The above results all assume the existence of a loop prolongation that fulfills the axioms of Definition \ref{def2.2.3}. Eilenberg and MacLane \cite{em} have proved  the   following 
theorem which ensures this existence:
{\teor{{\rm{[Eilenberg-MacLane]}} Let $G$ be a group, $K$ an Abelian group and $\sigma(g)$ an automorphism on $K$ for every $g\in G$. Then, given an $n$-cochain $\omega_n\in C_\sigma^{n}(G,K)$, 
there exists a prolongation $(L,\chi)$ of $G$ by $K$ and a system of representatives $\tau(g)$ such that $\omega^\tau_n=\omega_n$. Furthermore, the prolongation $(L,\chi)$ can be so 
constructed that the kernel of $\chi$ is Abelian.}\label{theorem2.2.1}}

This summary provides the general algebraic framework for the cocycle representations 
of the Galilean line group studied in section \ref{sec4}.  For more details, see references \cite{em,em1,em2, ai}.

\newpage

\end{document}